\documentclass[10pt]{article} 

\usepackage[utf8]{inputenc} 

\usepackage{geometry} 
\usepackage{graphicx} 

\usepackage{booktabs} 
\usepackage{array} 
\usepackage{paralist} 
\usepackage{verbatim} 
\usepackage{subfig} 
\usepackage{graphicx} 
\usepackage{fancyvrb } 
\usepackage{rotating} 
\usepackage{amssymb,amsmath} 

\usepackage{fancyhdr} 
\usepackage{sectsty}
\allsectionsfont{\sffamily\mdseries\upshape} 

\usepackage[nottoc,notlof,notlot]{tocbibind} 
\usepackage[titles,subfigure]{tocloft} 

\title{\textbf{\large Alternate Forms of the T-Matrix in Quantum State
Tomography}}
\author{\small Ramesh Bhandari\footnote{rbhandari@lps.umd.edu}\\[-1.0ex]
\small \textit {Laboratory for  Physical Sciences, 8050 Greenmead Drive, College
Park, Maryland 20740}}
\date{} 

\begin{document}
\maketitle

\begin{abstract}
In this paper, we focus on alternate forms of the T-matrix used in the
Maximum Likelihood Estimate (MLE) procedure for fitting the experimental data
collected in quantum state tomography experiments. In particular, we analyze
the single quantum state tomography case, deriving in the process three new
valid alternate forms for achieving optimality.
These alternative forms then serve as a consistency check, thus enhancing the
robustness of the MLE fitting process. One form, in particular, serves as a
useful compliment to the standard form normally employed. We subsequently
provide a generalization of these forms to the case of multiqubit state
tomography.
\end{abstract}
\thispagestyle{empty}


\section{\large Introduction}
A T-matrix is a theoretical construct in quantum state tomography to decipher
the quantum state from collected experimental data.  The data comprise
measurements of the Stokes parameters. The quantum state is expressed via a
density matrix, which is expressed in terms of the T-matrix and  then fitted to
the experimental data in an optimization process called the Maximum Likelihood
Estimate (MLE) [1]. The aim in the MLE process is to assign appropriate starting
values for the T-matrix parameters so that a global (and not a local) minimum is
achieved.  In an earlier paper [2], we analyzed the T-matrix traditionally used
in literature for single qubit state tomography and succeeded in providing 
novel expressions for the starting values of the T-matrix parameters based on
the measured experimental data. 
\\\\
In this note, we provide three  alternate forms of the T-matrix and report the results of our analysis, including the corresponding
sets of starting values for the different  T-matrices that can be used in the
MLE optimization technique. These forms then serve as a consistency check on the
fits to the density matrix. In particular, one of these alternate forms (called
Form B below)  compliments the standard form [1-3] (which we call Form A,
henceforth) in that where Form A fails, Form B works, and vice versa.  The other
forms (called Forms C and D below), while still being equally valid forms, are
not as easy to implement because the expressions for the starting values allow
for a larger region of the Stokes parameter space, where they may become
unstable. 
\\\\
While we focus on the T-matrix for single qubit tomography, we provide in 
Appendix A a generalization for multiqubit tomography. 
Two-photon quantum tomography has been employed in the past [1].
\section{\large Alternate Forms of the T-Matrix}
For the single qubit tomography, the  T-matrix used in literature [1-3] is 
\begin{equation}
T=
\left[
\begin{matrix}
t_1&0\\t_3+it_4&t_2
\end{matrix}
\right],
\end{equation}
and the density matrix $\rho$ is given by
\begin{equation}
\rho=\frac{T^\dagger T}{Tr(T^\dagger T)}.
\end{equation}
In order to derive alternate forms of the T-matrix, we start with the most
general representation of a 2x2 matrix. Consider 
\begin{equation}
T=
\left[
\begin{matrix}
a&c\\
d&b
\end{matrix}
\right],
\end{equation}
where parameters a,b,c, and d are complex in general; we alo assume in what
follows that these complex parameters are completely independent of each other.
This definition leads to 
\begin{equation}
\frac{T^\dagger T}{Tr(T^\dagger T)}
=
\frac{\left[
\begin{matrix}
|a|^2+|d|^2&a^*c+bd^*\\
ac^*+b^*d&|c|^2+|b|^2
\end{matrix}
\right]}
{|a|^2+|b|^2+|c|^2+|d|^2},
\end{equation}
which is Hermitian and has trace equal to 1,  by construction. Now any T-matrix
representation of a physical density matrix should contain 4 real independent
parameters. This is due to the fact that there are two angles $\theta$, $\phi$
of the Bloch sphere and a positive  "mixed" state parameter less than or equal
to 1, to provide a complete description of the quantum state under
consideration; a fourth parameter, which we call the "scaling" parameter, is
added to the T-matrix to facilitate numerical computation.  In order to generate
such a form of $\rho$, which is Hermitian and complex in general, two of the 4
real parameters must combine to form the complex off-diagonal elements of the 2
x 2 density matrix, i.e., one of the two terms in the expression, $ a^*c+bd^*$, 
of the off-diagonal element of the  matrix in Eq. 4  must be zero  in order to
accommodate the (at most) 4 real parameters that a physical density matrix is
constructed from. With these restrictions,  four structurally different forms
arise:
\\\\
A) \underline{c=0, with a and b real, and d being complex}
\\\\
Set $a=t_1, b=t_2, d=t_3+it_4$, which leads to
\begin{equation}
T=
\left[
\begin{matrix}
t_1&0\\
t_3+it_4&t_2
\end{matrix}
\right],
\end{equation}
and
\begin{equation}
\rho=\frac{T^\dagger T}{Tr(T^\dagger T)}\nonumber
\end{equation}
\begin{equation}
=\frac{
\left[
\begin{matrix}
t_1^2+t_3^2+t_4^2&t_2(t_3-it_4)\\
t_2(t_3+it_4)&t_2^2
\end{matrix}
\right]
}
{t_1^2+t_2^2+t_3^2+t_4^2},
\end{equation}
which is the standard from that has been employed in the past [1-3].
\\\\
B) \underline{d=0, with a and b real, and c being complex}
\\\\
Set $a=t_2, b=t_1, c=t_3+it_4$, which leads to
\begin{equation}
T=
\left[
\begin{matrix}
t_2&t_3+it_4\\
0&t_1
\end{matrix}
\right],
\end{equation}
and
\begin{equation}
\rho=\frac{T^\dagger T}{Tr(T^\dagger T)}\nonumber
\end{equation}
\begin{equation}
=\frac{
\left[
\begin{matrix}
t_2^2&t_2(t_3+it_4)\\
t_2(t_3-it_4)&t_1^2+t_3^2+t_4^2
\end{matrix}
\right]
}
{t_1^2+t_2^2+t_3^2+t_4^2},
\end{equation}
which is different from Form A (the standard form) that has been employed in the
past [1-3].
\\\\
C) \underline{a=0, with b and c real, and d being complex}
\\\\
Set $b=t_2, c=t_1, d=t_3+it_4$, which leads to
\begin{equation}
T=
\left[
\begin{matrix}
0&t_2\\
t_3+it_4&t_1
\end{matrix}
\right],
\end{equation}
and
\begin{equation}
\rho=\frac{T^\dagger T}{Tr(T^\dagger T)}\nonumber
\end{equation}
\begin{equation}
=\frac{
\left[
\begin{matrix}
t_3^2+t_4^2&t_2(t_3-it_4)\\
t_2(t_3+it_4)&t_1^2+t_2^2
\end{matrix}
\right]
}
{t_1^2+t_2^2+t_3^2+t_4^2},
\end{equation}
which is different from Forms A and B.
\\\\
D) \underline{b=0, with a and  d real, and c being complex}
\\\\
Set $a=t_2, d=t_1, c=t_3+it_4$, which leads to
\begin{equation}
T=
\left[
\begin{matrix}
t_1&t_3+it_4\\
t_2&0
\end{matrix}
\right],
\end{equation}
and
\begin{equation}
\rho=\frac{T^\dagger T}{Tr(T^\dagger T)}\nonumber
\end{equation}
\begin{equation}
=\frac{
\left[
\begin{matrix}
t_1^2+t_2^2&t_2(t_3+it_4)\\
t_2(t_3-it_4)&t_3^2+t_4^2
\end{matrix}
\right]
}
{t_1^2+t_2^2+t_3^2+t_4^2},
\end{equation}
which is different from Forms A, B, and C.  More forms can be derived by
interchanging the parameters $t_1$ and $t_2$, interchanging $t_3$ and $t_4$,
replacing $t_1$ by -$t_1$, and so forth, but these are not fundamentally
different.
 Do the three additional forms provide any advantage in the MLE process?
 We have analyzed the above three additional forms. Recalling that the
experimental density matrix is expressed in terms of the normalized Stokes
parameters:
\begin{equation}
\rho=\frac{1}{2}
\left[
\begin{matrix}
1+s_3&s_1-is_2\\s_1+is_2&1-s_3
\end{matrix}
\right].
\end{equation}
with the requirement
\begin{equation}
s_1^2+s_2^2+s_3^2 \leq 1,
\end{equation}
where the equality sign holds when the density matrix $\rho$ describes a
completely pure state,  we give the major results, including the critical 
starting values for the T-matrix parameters in the MLE optimization process;
these are expressed in terms of the experimentally determined Stokes
parameters. 
\section{\large Analysis of the Alternate Forms}
We now report the results of our analysis of these various alternate forms,
providing the sets of starting values for each form in the MLE search.
\subsection{\normalsize Form B}
1) det($\rho$)=0 leads to $t_2^2t_1^2\geq 0$, which implies 
\begin{equation}
\rho=
\left[
\begin{matrix}
0&0\\0&1
\end{matrix}
\right]
\end{equation}
when $t_2=0$. $t_1=0$ leads to a general pure state, as in Form A [2].
\\\\
2) Comparing Eq. 13 with Eq. 8 then leads to
\begin{equation}
t_1^2=\frac{(1-s_3^2-s_1^2-s_2^2)}{(1+s_3)^2}t_2^2,
\end{equation}
\begin{equation}
t_3=(\frac{s_1}{1+s_3})t_2,
\end{equation}
\begin{equation}
t_4=-(\frac{s_2}{1+s_3})t_2 .
\end{equation}
Eqs. 16, 17, and 18 give expressions for $t_1, t_3$, and $t_4$ in terms of the
Stokes parameters, $s_1, s_2$, and $s_3$, and the parameter $t_2$. With $t_2$ ,
say, fixed at 1, they serve as the starting values in the search for the minimum
in the MLE process. When $s_3$ is observed to be close to -1, one may then set
$t_2=0$ (see Eq. 15), with the remaining  parameters $t_1, t_3$, and $t_4$
initialized at some arbitrary values chosen  to be 1 each, for example. 
\\\\
\underline{Further Remarks}
\\\\
Form B is complimentary to Form A in the sense that if one fits the data
initially with Form A, then Form B serves as a back-up when the observed value
of $s_3$ is close to 1; in this situation, the expressions for the starting
values based on Form A (as we saw in [2]) become unstable near $s_3=1$, so it is
prudent to apply Form B whose starting values are stable near $s_3=1$.
Similarly, if one chooses to employ Form B in the MLE process, then Form A
serves as a backup in the vicinity of $s_3=1$.  
\\\\
For completeness, we analyze Forms C and D, and provide expressions of the
starting values based on these forms.
\subsection{\normalsize Form C}
1) det($\rho$)=0 leads to $(t_3^2+t_4^2)t_1^2\geq 0$, which implies 
\begin{equation}
\rho=
\left[
\begin{matrix}
0&0\\0&1
\end{matrix}
\right]
\end{equation}
when $t_3=t_4=0$. $t_1=0$ leads to a general pure state.
\\\\
2) Comparing Eq. 13 with Eq. 10 then leads to
\begin{equation}
t_1^2=\frac{(1-s_3^2-s_1^2-s_2^2)}{s_1^2+s_2^2}t_2^2,
\end{equation}
\begin{equation}
t_3=\frac{(1+s_3)s_1}{s_1^2+s_2^2}t_2,
\end{equation}
\begin{equation}
t_4=\frac{(1+s_3)s_2}{s_1^2+s_2^2}t_2 .
\end{equation}
Eqs. 20, 21, and 22 give expressions for $t_1, t_3$, and $t_4$ in terms of the
Stokes parameters, $s_1, s_2$, and $s_3$, and the parameter $t_2$. With $t_2$ ,
say, fixed at 1, they serve as the starting values in the search for the minimum
in the MLE process. Clearly, when $s_1\approx s_2\approx 0$, the above
expressions become unstable. Alternate expressions for the starting values of
the t parameters must be obtained for such situations; these are likley to be
more involved than in the case of Forms B and C. 
\subsection{\normalsize Form D}
1) det($\rho$)=0 leads to $(t_3^2+t_4^2)t_1^2\geq 0$, which implies 
\begin{equation}
\rho=
\left[
\begin{matrix}
0&0\\0&1
\end{matrix}
\right]
\end{equation}
when $t_3=t_4=0$. $t_1=0$ leads to a general pure state.
\\\\
2) Comparing Eq. 13 with Eq. 12 then leads to
\begin{equation}
t_1^2=\frac{(1-s_3^2-s_1^2-s_2^2)}{s_1^2+s_2^2}t_2^2,
\end{equation}
\begin{equation}
t_3=\frac{(1-s_3)s_1}{s_1^2+s_2^2}t_2,
\end{equation}
\begin{equation}
t_4=-\frac{(1-s_3)s_2}{s_1^2+s_2^2}t_2 .
\end{equation}
 Eqs. 24, 25, and 26 give expressions for $t_1, t_3$, and $t_4$ in terms of the
Stokes parameters, $s_1, s_2$, and $s_3$, and the parameter $t_2$. The
instability case near$s_1\approx s_2\approx 0$ is similar to Form C above.
\section{\large Summary}
We have derived four forms of the T-matrix, including the standard form used
currently in literature, in the construction of a physical density matrix for
the purposes of fitting the experimental data in the MLE optimization procedure.
This MLE procedure is often used in quantum state tomography. The three new
alternate forms, Forms B, C, and D, can be used for consistency checks. For all
the four forms of the T-matrix, the starting values derived from knowledge of
experimental data become unstable in some region of the space spanned by the
Stokes parameters. In particular, we show that Form B acts as a robust backup to
Form A, the standard form, when the latter becomes unstable. While Forms C and D
are equally valid forms to choose from, their implementation is relatively more
involved due to a larger region of instability, compared to Forms A and B. Our
recommendation therefore is to use Form A or Form B in the analysis of single
qubit state tomography, with the other acting as a backup. In Appendix A, we
give the generalization of Form B to multiple qubit state cases. The
generalization to Form A was given in Ref. [1].

\appendix
\numberwithin{equation}{section}
\section{\large Generalization of T-Matrix to a Multiqubit Case}
A T matrix for n qubits requires $2^{2n}$ independent parameters [1]. A general
form for Form A, the standard form , was provided in Ref.[1]. Here we provide a
generalization of the new Form B given in Eq. 7:
\begin{equation}
T=
\left[
\begin{matrix}
t_{2^n}&t_{2^n+1}+it_{2^n+2}&....&t_{4^n-1}+it_{4^n}\\
0&t_{2^n-1}&...&t_{4^n-3}+it_{4^n-2}\\
...&...&...&...\\
0&0&...&t_1
\end{matrix}
\right].
\end{equation}

\subsection{n=1; the single qubit case}
\begin{equation}
T=
\left[
\begin{matrix}
t_2&t_3+it_4\\
0&t_1
\end{matrix}
\right].
\end{equation}
\subsection{n=2; the two qubit case}
\begin{equation}
T=
\left[
\begin{matrix}
t_4&t_5+it_6&t_{11}+it_{12}&t_{15}+it_{16}\\
0&t_3&t_7+it_8&t_{13}+it_{14}\\
0&0&t_2&t_9+it_{10}\\
0&0&0&t_1.
\end{matrix}
\right].
\end{equation}
Similarly, Form C (Eq. 10) and Form D (Eq. 12) can be generalized.


\end{document}